# Simulating Entanglement in Classical Computing

Subhash Kak

*Abstract:* This note shows how quantum entanglement may be simulated in classical computing. The simulated entanglement protocol is implemented using oblivious transfer in the simplest case and other many-to-one mappings in more general cases. For the case where the mapping is of order *k,* we prove a theorem that gives us the function of the entangled state. A useful generalization beyond this research will be the implementation of quantum states with arbitrary probability amplitude functions.

**Introduction**
Entanglement is a quantum mechanical concept that describes the correlation between the properties of two or more quantum objects even when they are far apart [1]. Although, before measurement the outcome of the interaction of the equipment with these objects is unpredictable, after one of them has been measured the measurement of the second one is fully determined. Consider, for example, two individuals, Alice and Bob, who share photons that form the entangled pair $\frac{1}{\sqrt{2}}(|HH\rangle + |VV\rangle)$. In the measurement of each of these photons the probability of getting horizontally or vertically polarized photons is the same. If Alice finds that her photon is H (or V), so is the photon of Bob. Of course, if the entanglement is not maximal then the measurement of one does not fully provide knowledge of the other.

Entanglement in quantum system is a consequence of the laws of quantum mechanics and it has found a role in quantum cryptography protocols [2],[3]. It would be nice, therefore, to simulate entanglement in classical computing. Since maximally entangled states are a subset of arbitrary quantum states, the larger problem is that of the implementation of such arbitrary states.

To simulate entanglement in classical computing one would require a scheme that leaves the outcome unknowable (to all parties) until the measurement is made and still allows the same (or complementary) result to show up at the remote node. Since entanglement can be in different degrees, one would also need to be able to associate the second measurement with different values of probability. Thus the measurement at the first node need not, in the general case, completely determine the measurement at the second node. We assume that in simulated entanglement the underlying complex probability amplitudes of the quantum state are not a concern although that may very well be considered by the researcher in a separate study.

Here we show how the simulation of entanglement can be achieved by the use of the basic oblivious transfer mapping [4], and cubic [5], quadratic and higher-degree mappings. Considering that the mapping is of order k, which is an odd number, we prove a theorem that gives the function of the entangled state.

We assume that there are two parties, Alice and Bob, who either are able to find a specific secret, or not. Let the event that one of them finds the secret be represented by Y (for "yes") and the event when the secret is not found to be N (for "no"). In general, we seek a protocol for the realization of the joint state:

$$\sqrt{a}|YY\rangle + \sqrt{b}|YN\rangle + \sqrt{c}|NY\rangle + \sqrt{d}|NN\rangle$$

where $a+b+c+d = 1$.

In the maximally entangled state, $a = d = 1/2$ and $b = c = 0$. In this paper we present a solution for the maximally entangled case as well as a variety of specific cases where $b=c \neq 0$ (the equality of $b$ and $c$ is a consequence of the symmetry of structure assumed in this paper). Further research should seek implementation of quantum states with arbitrary values of probability amplitudes.

**Simulated Entanglement Protocol (SEP)**
The basic simulated entanglement protocol, which is a modification of the oblivious transfer protocol, is described in Figure 1.

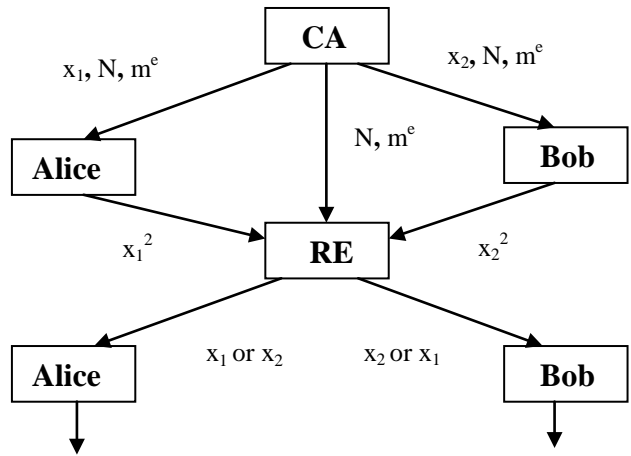

**Figure 1.** The SE protocol where $x_1^2 = x_2^2 \bmod N$ and $x_2 \neq N - x_1$

The system consists of two special agents: CA (Certification Authority) and RE (Root Extractor). CA begins a communication process with a secret and the objective is for the two parties, Alice and Bob, to obtain the secret in a complementary fashion based on the computations by RE, but neither CA nor RE should know which one of the two has obtained the secret. The RE can be an application that performs its computation according to an algorithm that is not known to any of the three parties CA, Alice, or Bob. The question of how probability might be viewed from the perspective of physical theory (as in [6]-[10]) in the workings of RE is beyond the scope of this paper.

The secret is the message m, which is raised to an exponent e modulo a composite number N, the roots of which are only known to CA and RE; e and N are publicly announced numbers. To determine the secret it is essential for Alice and Bob to have the root of $x^2$ that they do not already possess. Since RE does not know which roots were sent by CA to Alice and Bob, his choice determines which one of the two will obtain the secret. Either both Alice and Bob will have the secret or neither of them will have it.

The protocol of Figure 1, therefore, simulates the maximally entangled quantum state:

$$\sqrt{\frac{1}{2}}|YY\rangle + e^{i\theta}\sqrt{\frac{1}{2}}|NN\rangle$$

for arbitrary $\theta$ as prob($|YY\rangle$) = prob($|NN\rangle$) = 1/2.

**SEP with probability 2/3**

Entanglement in the quantum world does not always occur with the same probability. For example, a quantum pair of objects may be entangled to the extent as given by $\frac{1}{\sqrt{3}}|HH\rangle + \sqrt{\frac{2}{3}}|VV\rangle$, which means that horizontally polarized photons are obtained with probability of 1/3 and vertically polarized photons are obtained with a probability of 2/3. As in the previous case, such a simulation will not be able to distinguish between quantum states such as $\frac{1}{\sqrt{3}}|HH\rangle + \sqrt{\frac{2}{3}}|VV\rangle$ and $\frac{1}{\sqrt{3}}|HH\rangle + e^{i\theta}\sqrt{\frac{2}{3}}|VV\rangle$ for different values of $\theta$. For simplicity of presentation, we will not explicitly mention the angle $\theta$ further.

Simulating such a situation requires that probabilities other than ½ of the oblivious transfer protocol be realized. In general, this requires that RE should compute higher roots of the number sent to him and then he should pick random subsets of these numbers and send them back to Alice and Bob. This can be done using the scheme of higher order mappings [5].

**Example 1**. Consider probabilities of 2/3 and 1/3 that are associated with the measurement of the states *Y* and *N*. This would correspond to the quantum state $\sqrt{a}|YY\rangle + \sqrt{b}|YN\rangle + \sqrt{c}|NY\rangle + \sqrt{d}|NN\rangle$ where $a+b$ = 2/3 and $c+d$ = 1/3 and, clearly, this would have many different solutions. We can realize it for a special case by the cubic mapping as shown below.

Consider N = 77. Since φ(77)=60 which is divisible by 3, numbers that have cubic roots will have three of those. For example the three roots of the following numbers that are co-prime with 77 are given in the table below:

Table 1. Cubic roots of n mod 77

| n=$x_i^3$ mod 77 | $x_1$ | $x_2$ | $x_3$ |
|---|---|---|---|
| 1 | 1 | 23 | 67 |
| 6 | 19 | 41 | 52 |
| 8 | 2 | 46 | 57 |
| 13 | 40 | 62 | 73 |
| 15 | 16 | 60 | 71 |
| 20 | 26 | 48 | 59 |
| 27 | 3 | 47 | 69 |
| 29 | 39 | 50 | 72 |
| 34 | 12 | 34 | 45 |
| 36 | 9 | 53 | 64 |
| 41 | 13 | 24 | 68 |
| 43 | 32 | 43 | 65 |
| 48 | 5 | 27 | 38 |
| 50 | 8 | 30 | 74 |
| 57 | 18 | 29 | 51 |
| 62 | 6 | 17 | 61 |
| 64 | 4 | 15 | 37 |
| 69 | 20 | 31 | 75 |
| 71 | 25 | 36 | 58 |
| 76 | 10 | 54 | 76 |

We propose the use of the protocol of Figure 2.

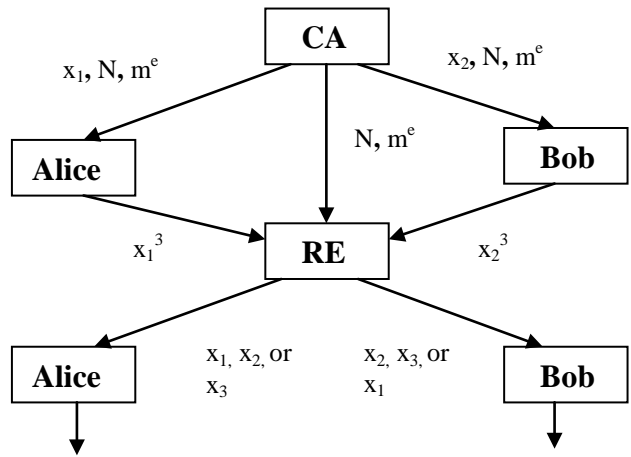

**Figure 2.** The SE protocol for non-equal probabilities

*Step 1.* The Certification Authority (CA) sends two different cubic roots, $x_1$ and $x_2$, of some chosen number, x, to Alice and Bob. Alice and Bob find the cube of this number modulo N and send it to RE. This number is $x = x_1^3 \mod N = x_2^3 \mod N$.

*Step 2.* RE knows the factors of N and, therefore, he is able to find the three cubic roots of x, which are $x_1, x_2, x_3$. He now picks two of these roots and sends them to Alice and Bob, respectively.

*Step 3.* Alice and Bob are able to or unable to find the factors of N based on what they have received. Each one of them is able to find the factors if the root received is different from the one they started with.

The probability that each one of them will find the factors is 2/3. This is seen by the table below that summarizes the 6 different cases associated with the choices available to them.

Table 3. Cases in the cubic mapping

| Alice's number | Bob's number | Alice's ability to factor | Bob's ability to factor |
|---|---|---|---|
| $x_1$ | $x_2$ | no | no |
| $x_1$ | $x_3$ | no | yes |
| $x_2$ | $x_3$ | yes | yes |
| $x_2$ | $x_1$ | yes | yes |
| $x_3$ | $x_1$ | yes | yes |
| $x_3$ | $x_2$ | yes | no |

For example, if Alice and Bob start with cubic roots 5 and 38 of the number 48. Each of them sends 48 to RE who sends back one of the three 5, 27, and 38 each to Alice and Bob. If Alice receives 27 or 38, all she needs to do is to subtract her own number 5 from the numbers received, yielding 22 and 33, respectively. The g.c.d. of any of these numbers and N=77 is one of the factors of 77.

The entangled state that is realized by the example of Table 3 is therefore:

$$\sqrt{\frac{1}{2}}|YY\rangle + \sqrt{\frac{1}{6}}|YN\rangle + \sqrt{\frac{1}{6}}|NY\rangle + \sqrt{\frac{1}{6}}|NN\rangle$$

We conclude that we have not quite realized the state $\sqrt{\frac{2}{3}}|YY\rangle + \sqrt{\frac{1}{3}}|NN\rangle$ that we set out to implement although the cumulative probabilities of Y and N are 2/3 and 1/3, respectively.

**SEP with other probabilities**

SEP with probability of $\frac{k-1}{k}$, that is simulation of the quantum state $\sqrt{a}|YY\rangle + \sqrt{b}|YN\rangle + \sqrt{c}|NY\rangle + \sqrt{d}|NN\rangle$, where a+b = $\frac{k-1}{k}$ and c+d = $\frac{1}{k}$ can be easily implemented by the use of a *k*th degree root extractor in a protocol similar to the one in Figure 2.

Consider, for example, *k*=4. We can use operations with respect to N=77 as φ(77) is divisible by 4. Thus the four roots of 4 are 3, 25, 52, and 74. Likewise, the four roots of 16 are 2, 9, 68, and 75. But this case is essentially the same as taking square roots in the Rabin protocol [4].

**Example 2.** Consider, *k*=5. We can still use N=77 as φ(77) is also divisible by 5. The 5 roots of 32 are 2, 30, 51, 65, and 72, and the five roots of 67 are 9, 16, 23, 37, and 58. The state represented in each of the two examples will be:

$$\sqrt{\frac{13}{20}}|YY\rangle + \sqrt{\frac{3}{20}}|YN\rangle + \sqrt{\frac{3}{20}}|NY\rangle + \sqrt{\frac{1}{20}}|NN\rangle$$

**Theorem**. For general odd k, a generalization of Protocol 2 will simulate the entangled state:

$$\sqrt{\frac{k^2-3k+3}{k(k-1)}}|YY\rangle + \sqrt{\frac{k-2}{k(k-1)}}|YN\rangle + \sqrt{\frac{k-2}{k(k-1)}}|NY\rangle + \sqrt{\frac{1}{k(k-1)}}|NN\rangle$$

*Proof.* The total number of cases to consider for the distribution of the roots to Alice and Bob by RE is $k(k-1)$. Of these cases precisely one leads to neither being able to find the secret (namely where both Alice and Bob receive what they had started out with). Of the remaining there are exactly $(k-2)$ cases where Alice cannot find the secret since the total number of negative cases for her equals $(k-1)$. By symmetry, there is an identical number of similar cases for Bob. Therefore, the number of cases out of $k(k-1)$ that both are able to find the secret is $k(k-1) - 2(k-2) - 1 = k^2 - 3k - 3$ which proves the result. ∎

Since $\frac{k^2-3k+3}{k(k-1)} + \frac{k-2}{k(k-1)} = \frac{k-1}{k}$, both Alice and Bob have the probability of $\frac{k-1}{k}$ to find the secret.

**Discussion**

There are two obvious research problems that are suggested by this research. The first of these is the design of protocols that implement the general state function $\sqrt{a}|YY\rangle + \sqrt{b}|YN\rangle + \sqrt{c}|NY\rangle + \sqrt{d}|NN\rangle$ for arbitrary *a, b, c, d*. The second problem is: Can simulated entanglement be used gainfully in a key distribution protocol?

**References**
1. E. Schrödinger, Discussion of probability relations between separated systems. Mathematical Proceedings of the Cambridge Philosophical Society 31: 555–563, 1935; 32: 446–451, 1936.
2. T. Jennewein, C. Simon, G. Weihs, H. Weinfurter, and A. Zeilinger, Quantum cryptography with entangled photons. Phys. Rev. Lett. 84: 4729-4732, 2000.
3. G. Gordon and G. Rigolin, Quantum cryptography using partially entangled states. Optics Communications 283: 184-188, 2010.
4. M.O. Rabin. Digitalized signatures and public-key functions as intractable as factorization. MIT/LCS/TR-212, MIT Laboratory for Computer Science, 1979.
5. S. Kak, The cubic public-key transformation. Circuits Systems Signal Processing 26: 353-359, 2007.
6. R.P. Feynman, Simulating physics with computers. Intl. J. Theoretical Physics 21: 467-488, 1982.
7. R. Landauer, The physical nature of information. Physics Letters A 217: 188-193, 1996.
8. S. Kak, Information, physics and computation. Foundations of Physics 26: 127-137, 1996.
9. S. Kak, The initialization problem in quantum computing. Foundations of Physics 29: 267-279, 1999.
10. D.K. Ferry, Quantum computing and probability. J. Phys.: Condens. Matter 21 474201, 2009.